\documentclass[twocolumn,showpacs,amsmath,amssymb]{revtex4}
\usepackage{graphicx}
\usepackage{amsmath}

\usepackage{color}

\begin{document}
\title{Far-from-equilibrium Sheared Colloidal Liquids: Disentangling Relaxation, Advection, and Shear-induced Diffusion}
%Far-from-equilibrium Sheared Colloidal Liquids: Double Scaling in the Crossover Regime}
%Far-from-equilibrium Sheared Colloidal Liquids: Two Saturations on One Universal Curve\\
%Double Scaling of Far-from-equilibrium Sheared Colloidal Liquids: Relaxation, Affine motion, and Shear-induced Diffusion}
\author{Neil Y.C. Lin$^1$, Sushmit Goyal$^2$, Xiang Cheng$^1,^3$, Roseanna N. Zia$^2$, Fernando A. Escobedo$^2$, Itai Cohen$^1$}
\affiliation{$^1$Department of Physics, $^2$Department of Chemical and Biomolecular Engineering,Cornell University, Ithaca, New York 14853}
%\affiliation{ Cornell University, Ithaca, New York 14853}
\affiliation{$^3$Department of Chemical Engineering and Materials Science, University of Minnesota, Minneapolis, Minnesota 55455}
\date{\today}

\begin{abstract}
Using high-speed confocal microscopy, we measure the particle positions in a colloidal suspension under large amplitude oscillatory shear.
Using the particle positions we quantify the \textit{in situ} anisotropy of the pair-correlation function -- a measure of the Brownian stress. From these data, we find two distinct types of responses as the system crosses over from equilibrium to far-from-equilibrium states.
The first is a nonlinear amplitude saturation that arises from shear-induced advection, while the second is a linear frequency saturation due to competition between suspension relaxation and shear rate.
In spite of their different underlying mechanisms, we show that all the data can be scaled onto a master curve that spans the equilibrium and far-from-equilibrium regimes, linking small amplitude oscillatory to continuous shear. This observation illustrates a colloidal analog of the Cox-Merz rule and its microscopic underpinning. Brownian Dynamics simulations show that interparticle interactions are sufficient for generating both experimentally observed saturations. %Furthermore, the simulation also shows that Cox-Merz rule can generally apply to other far-from-equilibrium systems as long as three essential ingredients are present: relaxation, advection, and shear-induced diffusion.
 
\end{abstract}


\maketitle
%%%%%%%%%%%%%%%%%%%%%%%%%%%%%%%%%%%%%%%%%%%%%%%%%%%%%%%%%%%%%%%%%%%%%%%%%%%%%%%%
%%%%%%%%%%%%%%%%%%%%%%%%%%%%%%%%%%%%%%%%%%%%%%%%%%%%%%%%%%%%%%%%%%%%%%%%%%%%%%%%

\section{Introduction}

While ubiquitous across many time and length scales, far-from-equilibrium behavior is still largely uncharted territory\cite{Jaeger2013}.
To understand such systems, the crossover from nearly equilibrium to far-from-equilibrium regimes provides crucial insights that bridge distinct concepts developed in either limit.
Furthermore, this poorly explored crossover is important for understanding natural phenomena such as nonlinear elasticity\cite{Storm2005}, flow-induced rejuvenation\cite{Bonn2002, Derec2003}, and shear thinning\cite{Wagner2009, Evans,Brader}, which all occur in industrial settings. A particularly fascinating and relevant example is the nonlinear stress response of fluids under large amplitude oscillatory shear (LAOS)\cite{Hyun2011,WagnerLAOS}. 
By varying the amplitude and frequency separately, LAOS disentangles the underlying dynamics that are usually convolved in such far-from-equilibrium systems.
 Despite great efforts, conventional flow measurement techniques\cite{Narumi, Hyun2002, Brader2010, Ewoldt2009, Giuseppe2012} have had difficulties elucidating the origins of these nonlinear behaviors without information about the fluid microstructure.

Because of their experimentally accessible time and length scales, hard-sphere colloids are an ideal model system to study nonlinear behaviors in far-from-equilibrium systems\cite{Weeks2000, WillemK.Kegel2000, Toyabe}. 
Here, by mounting a custom shear cell on a confocal microscope, we directly image colloidal liquids and quantify the suspension structure using the pair correlation function $g(\vec r)$. 
As the suspension is sheared, distortions of $g(\vec r)$ increase and lead to the Brownian stresses that arise from the thermal motion of particles\cite{Brady1993,Foss2000}. We quantify this $g(\vec r)$ change to capture the \textit{suspension structure response} in the crossover regime bounded by lightly perturbed states (nearly-equilibrium) and strongly driven states (far-from-equilibrium). This approach circumvents many difficulties encountered by conventional flow characterization techniques because it identifies the microscopic origin of the  stress response. In contrast to the previously explored linear response regime\cite{Cheng2011}, in this report we focus on the nonlinear response by performing LAOS measurements. Our data shows two distinct structure responses that collapse onto a master curve revealing the interplay between thermal relaxation, advection, and shear induced diffusion in the crossover regime.

\begin{figure*} [htp]
\includegraphics[width=1 \textwidth]{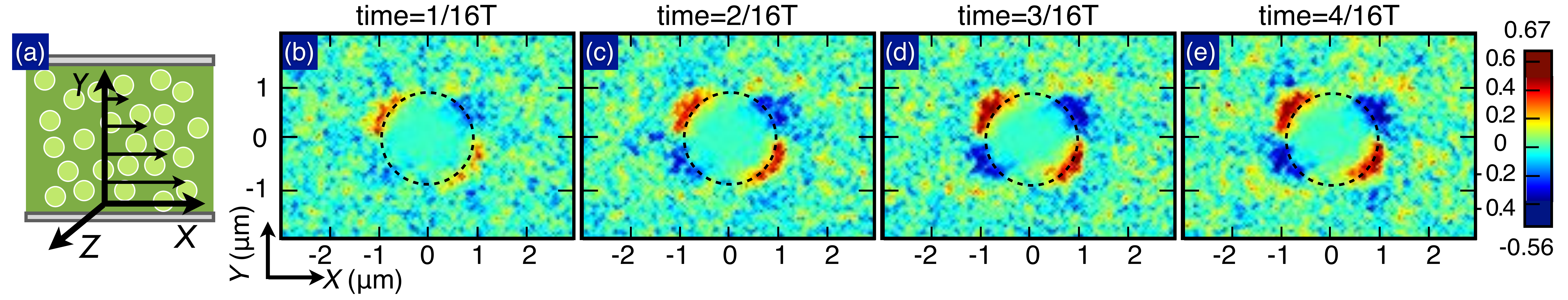}
\caption{(Color online) Shear configuration and coordinates (a), and $\hat X-\hat Y$ projection of $\Delta g(\vec{r})$ (b)-(e). The sheared suspension has volume fraction $\phi=0.28$. The values of $\omega$ and $\gamma_{0}$ are 0.13s$^{-1}$ and 3.34. Plots (b)-(e) show $\Delta g(\vec{r})$ for a quarter cycle of shear. (b)-(c), Anisotropy of $\Delta g(\vec{r})$ increases. (d)-(e), With increased displacement we find very little additional change in structure.}
\label{fig:fig1_v2}
\end{figure*}

\section{Experiment}

In our experiments we use silica particles with radius $a$ = 490 nm and 2$\%$ polydispersity, suspended in an index matching water-glycerine mixture whose viscosity $\eta_{0}$ = 0.06 Pa$\cdot$s. We add 1.25 mg/ml of fluorescein sodium salt to dye the solvent for imaging\cite{Cheng2011}. The electrostatic screening length is $\sim$10 nm, hence particle interactions are nearly hard-sphere, see Appendix A. Experiments are conducted on six samples with volume fractions $0.17\leq \phi \leq0.44$. 

To image the suspension structure during shear, we mount a piezoelectrically driven parallel plate shear cell on a fast scanning confocal microscope. The cell consists of a moveable cover slip as the bottom plate and a silicon wafer as the top plate (Fig.~\ref{fig:fig1_v2} (a)). We fix the separation along the gradient or $\hat Y$ direction to be 6.5 $\pm$ 0.2 $\rm{\mu m}$ with both plates aligned to within 0.0075$^\circ$ of one another by adjusting set screws. A solvent trap is used to prevent evaporation of the solvent. We sinusoidally shear the suspension over a range of amplitudes $0.06 \leq \gamma_{0} \leq 3.34$ and angular frequencies $ 0.006$ s$^{-1} \leq \omega \leq 0.628$ s$^{-1}$, where $\gamma_0$ is the shear strain amplitude that characterizes the ratio of the shear plate displacement to gap size. Capturing 216 frames per second, we acquire stacks of 40 images in less than 0.2 s. This scan rate is one hundred times faster than the highest shear frequency used and hence avoids distortion or mismatch between images. Prior to measurement, the sample is sheared at $\gamma_{0}=3.34$ with $\omega$=12.56 s$^{-1}$ to generate consistent initial conditions and then sheared with the target $\gamma_0$ and $\omega$ for five minutes, or 10 times the system relaxation time, 30 s. No hysteresis is observed throughout the measurements. Since the relaxation time for the suspension to diffuse back from the reservoir surrounding the shear zone is on the order of hours, the conducted experiments do not last long enough for shear induced migration effects to be substantial. We also find that the shear velocity profile is linear. Finally, we do not observe any driven assembly and structure organization under shear.

\section{Results}

To measure the three dimensional structure, we image the suspension, determine the particle positions\cite{crocker}, and construct the three dimensional pair correlation function $g(\vec{r})$. This function is proportional to the probability of finding a particle at position $\vec{r}$ with respect to each particle center. 

We exclude the particles that are aligned by the shearing surfaces by reducing the measurement window in the $g(\vec r)$ calculation. This exclusion allows for accurately calculating the suspension structure from the bulk of the sample. The three dimensional $g(\vec{r})$ is used to quantify the anisotropy in the shear induced structure. Under shear, particles accumulate along the maximal compression axis (MCA) -- oriented at $45^{\circ}$ to the flow direction $\hat{X}$ -- resulting in a distorted $g(\vec{r})$ \cite{Brady1993, Foss2000, Brader, Lehigh, Roseanna, Koumakis, Vermant2005}. We illustrate this effect by plotting $\Delta g = g_{xy}(t)-g_{xy}(t=0)$ over a quarter of an oscillation cycle, characterized by a large strain amplitude $\gamma_0=3.34$ and high frequency $\omega = 0.126$ s$^{-1}$. Here, $g_{xy}$ is a two dimensional cross section of $g(\vec{r})$ centered at $z=0$ with a width of 1.4 $\mu$m, and the response cycle starts at time $t=0$ where $g_{xy}$ is isotropic (Fig.~\ref{fig:fig1_v2} (b)-(e)). At this large strain amplitude and frequency, $\Delta g$ strongly saturates after the first eighth of the period $T$, as indicated by the negligible difference between Fig.~\ref{fig:fig1_v2}(d) and (e). In addition, we find that at large strains, particle accumulation tends to be broad and extends to angles below $45^{\circ}$. The unsubtracted $g_{xy}(t)$ is shown in Appendix B.

To quantify these observations, the structure signature 
\begin{equation}
\Psi(t) = \left\langle \oint g(\vec{r},t) \hat{X} \hat{Y} d\Omega \right \rangle, %\propto \langle \sigma_B \rangle
\label{eq:model}
\end{equation}

is defined, where $\hat{X}$ and $\hat{Y}$ are the unit vectors defined in Fig.~\ref{fig:fig1_v2}(a), $d\Omega$ is the solid angle, and the bracket denotes averaging over the interval $1.84a\leq r\leq 2.35a$. $\Psi$ is zero when the particle configuration is isotropic, and increases in value as particles line up along the MCA. Imaging artifacts associated with particle featuring errors broaden the first peak of $g(\vec r)$. To account for this effect, the lower bound of the averaging interval is chosen to be 80 nm smaller than $2a$, while the upper bound is chosen to be at the first peak of $g(\vec r)$ where $r=2.35a$ (See Appendix C).

\begin{figure} 
\includegraphics[width=0.5\textwidth]{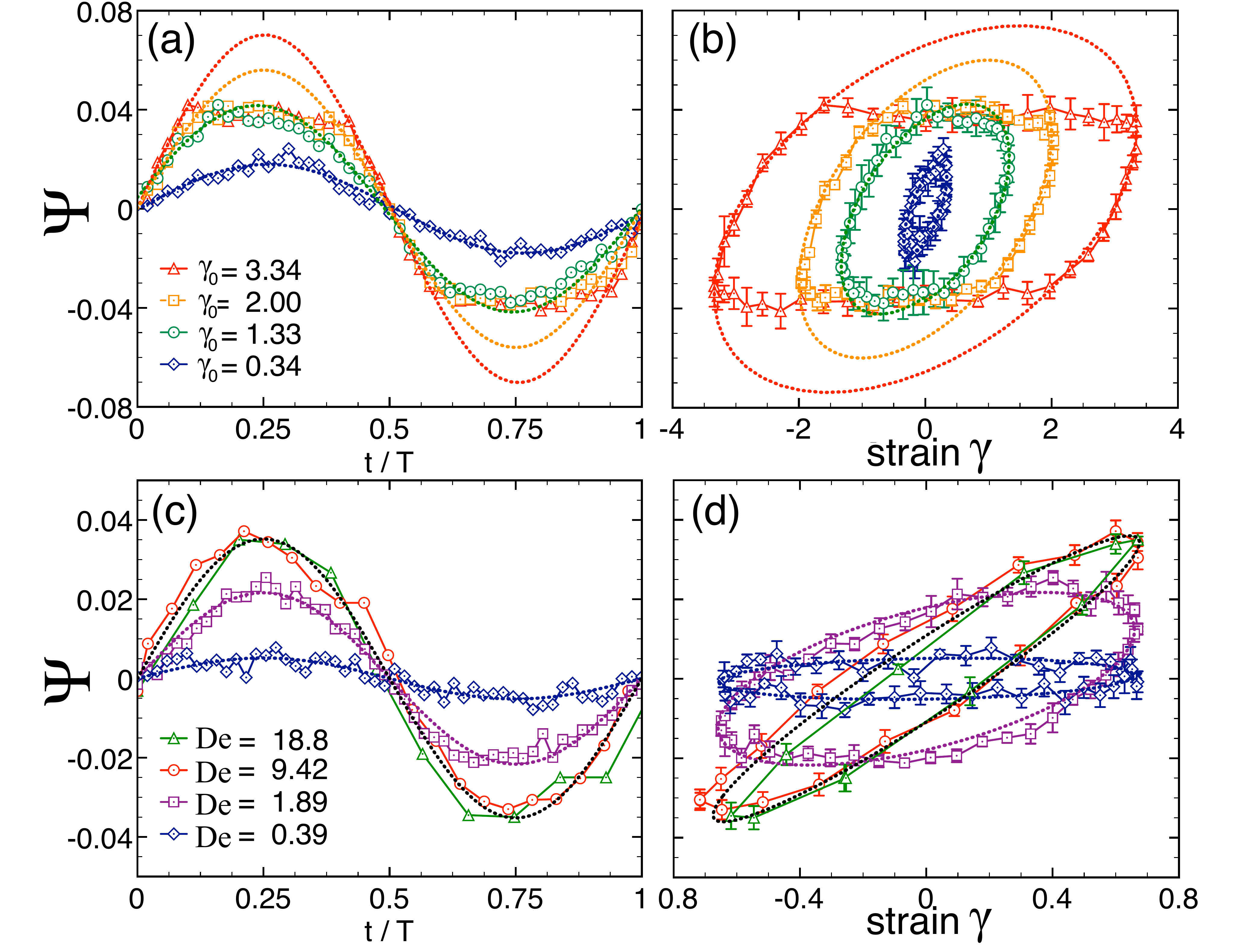}
\caption{(Color online) Structure response versus normalized time and Lissajous-Bowditch curves. (a), Four $\Psi(t)$ curves averaged over five measurements for De $=3.78$ and $0.34\leq \gamma_0 \leq 3.34$. Dashed lines are the sinusoidal fit to the data in the linear regime where $| \Psi(t) |< 0.03$. (b), The L-B curves for the panel (a) dataset. (c), Four different curves for $\gamma_0 = 0.67$ and $0.39<$ De $<18.8$ versus the normalized time. The black dashed line is the best fit to the combined data for De $=9.42$ and 18.8. (d), The L-B curves of the data set from panel (c). Error bars correspond to the standard error of $\Psi(t)$.}
\label{fig:result2}
\end{figure}

We explore the structural saturation (Fig.~\ref{fig:fig1_v2} (d) and (e)) by comparing $\Psi (t)$ at four strain amplitudes (Fig.~\ref{fig:result2} (a) and (b)). The dimensionless frequency, or Deborah number is De $=6 \pi a^3 \omega \eta_0 /k_B T$, and fixed at De $=3.78$. Physically, De is the ratio of oscillation frequency to suspension relaxation rate, while $\gamma_0$De defines the P$\rm \acute{e}$clet number -- the ratio of advective forcing to entropic restoring force. We explore the saturation at high frequencies by measuring $\Psi (t)$ at fixed strain amplitude $\gamma_0 = 0.67$, and at four values of De (Fig.~\ref{fig:result2}(c) and (d)). Remarkably, the observed saturation of $\Psi(t)$ is qualitatively different depending on whether we increase $\gamma_0$ or De. 

For the constant-De data, at low strain amplitudes the structure response of the suspension is linear and well described by a sinusoid (dashed black line Fig.~\ref{fig:result2} (a)). At a strain of $\gamma_0 \geq 1.33$ we find that $\Psi (t)$ begins to saturate at its maximal values. This saturation is more pronounced at even higher strain amplitudes as illustrated by the $\gamma_0 = 2.00$ and 3.34 data. Thus, we find that $\Psi(t)$ deviates from a linear sinusoidal response at large $\gamma_0$. To demonstrate the phase relationship between $\Psi(t)$ and the instantaneous strain, we plot the corresponding Lissajous-Bowditch (L-B) curves in Fig.~\ref{fig:result2} (b). In this representation, a linear viscoelastic response is depicted by an ellipse whose orientation and enclosed area correspond to the material elasticity and viscous dissipation, respectively. For large $\gamma_0$ at fixed De, the elliptical linear response saturates at a $\gamma_0$-independent plateau. We plot the Pipkin diagram in Appendix D.

$\Psi(t)$ shows strikingly different behavior when we hold $\gamma_0$ at 0.67 and sweep over De. At small De we find the response is sinusoidal and purely viscous as illustrated by the horizontal orientation of the De $=0.39$ L-B curve (Fig.~\ref{fig:result2} (d)). As De increases to $1.89$, the L-B curve remains elliptical but acquires a significant tilt -- indicating a harmonic response with increased elasticity. As De increases further, the L-B curves overlap indicating a De-independent harmonic response (Fig.~\ref{fig:result2}(c) and (d)). 

\begin{figure} 
\includegraphics[width=0.48 \textwidth]{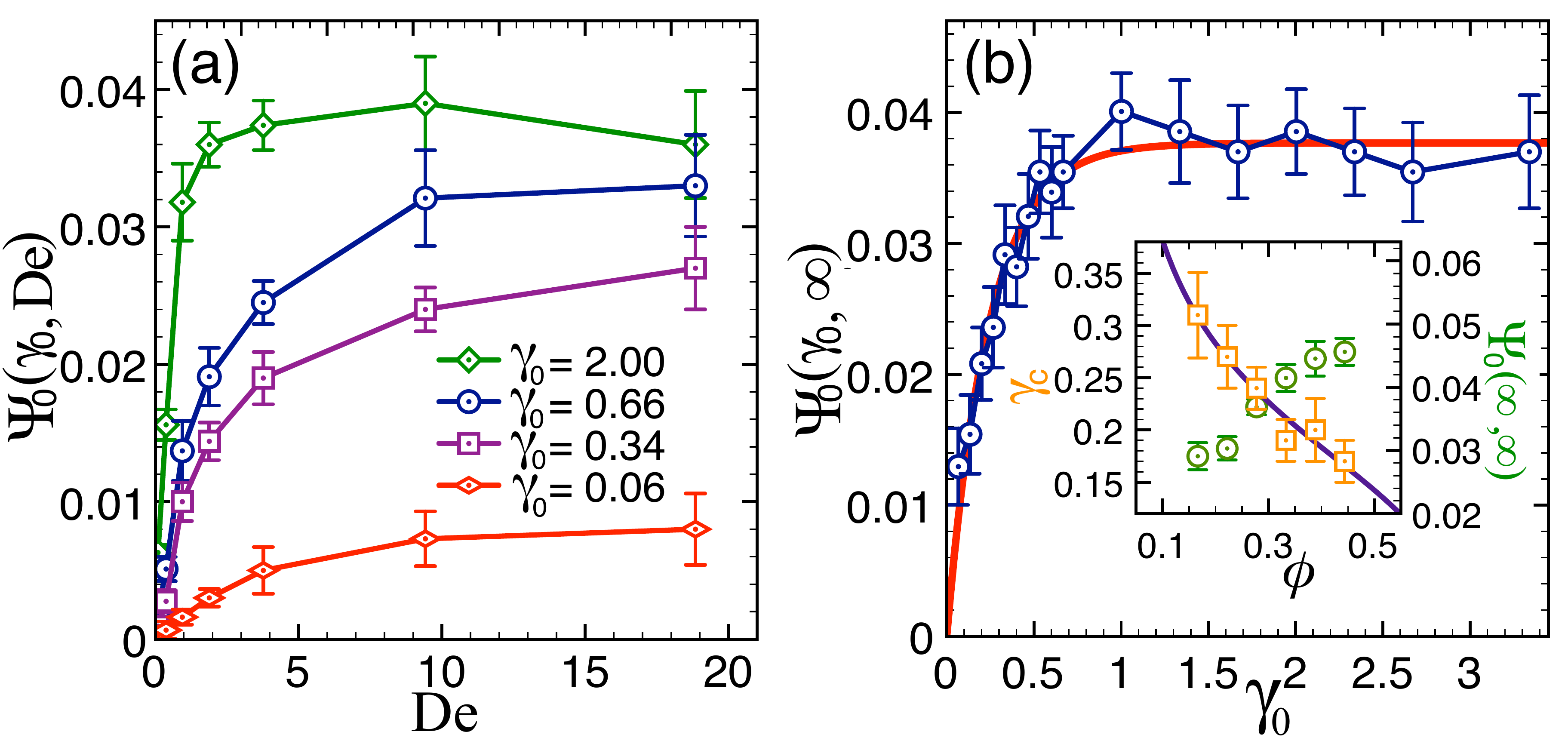}
\caption{(Color online) Structure response saturations. (a), Peak value of structure response in each cycle, $\Psi_0(\gamma_0,{\rm De})$ versus De with four representative amplitudes. Error bars depict the standard error of $\Psi_0(\gamma_0,{\rm De})$ over five cycles. (b), The saturation value $\Psi_0(\gamma_0,\infty)$, measured at De $=18.8$, is plotted versus $\gamma_0$. The red curve is a fit of $\Psi_0(\infty,\infty)(1 - e^{-\gamma_0/\gamma_c})$ to the data. The inset illustrates the volume fraction dependence of $\gamma_c$ and $\Psi_0(\infty,\infty)$. The solid line in the inset has the form  $\gamma_c \propto (\frac{0.64-\phi}{\phi})^{1/3}$. Data points are averaged over ten measurements and error bars depict standard errors.}
\label{fig:fig3}
\end{figure}

To understand the saturation of $\Psi(t)$ with increasing $\gamma_0$ and De, we track its maximum value per cycle, $\Psi_0(\gamma_0, \rm De)$. We plot $\Psi_0(\gamma_0, \rm De)$ versus De for four representative values of $\gamma_0$ in Fig.~\ref{fig:fig3}(a). For $\gamma_0 = 0.06$ we find $\Psi_0(\gamma_0, \rm De)$ increases linearly below De $\approx3$ and saturates to a plateau value $\Psi_0(\gamma_0,\infty)$ at large De. Similar trends are observed in all datasets but with different plateau values depending only on $\gamma_0$. To quantify changes in these plateau values, we plot $\Psi_0(\gamma_0,\infty)$ versus $\gamma_0$ in Fig.~\ref{fig:fig3} (b). We find that the data are well fit by an exponential saturation, $\Psi_0(\gamma_0,\infty)=\Psi_0(\infty,\infty)(1 - e^{-\gamma_0/\gamma_c})$ (red curve) indicating a linear growth at low $\gamma_0$ and saturation to $\Psi_0(\infty,\infty)$ beyond a cutoff strain amplitude $\gamma_c$. 

\begin{figure} 
\includegraphics[width=0.48 \textwidth]{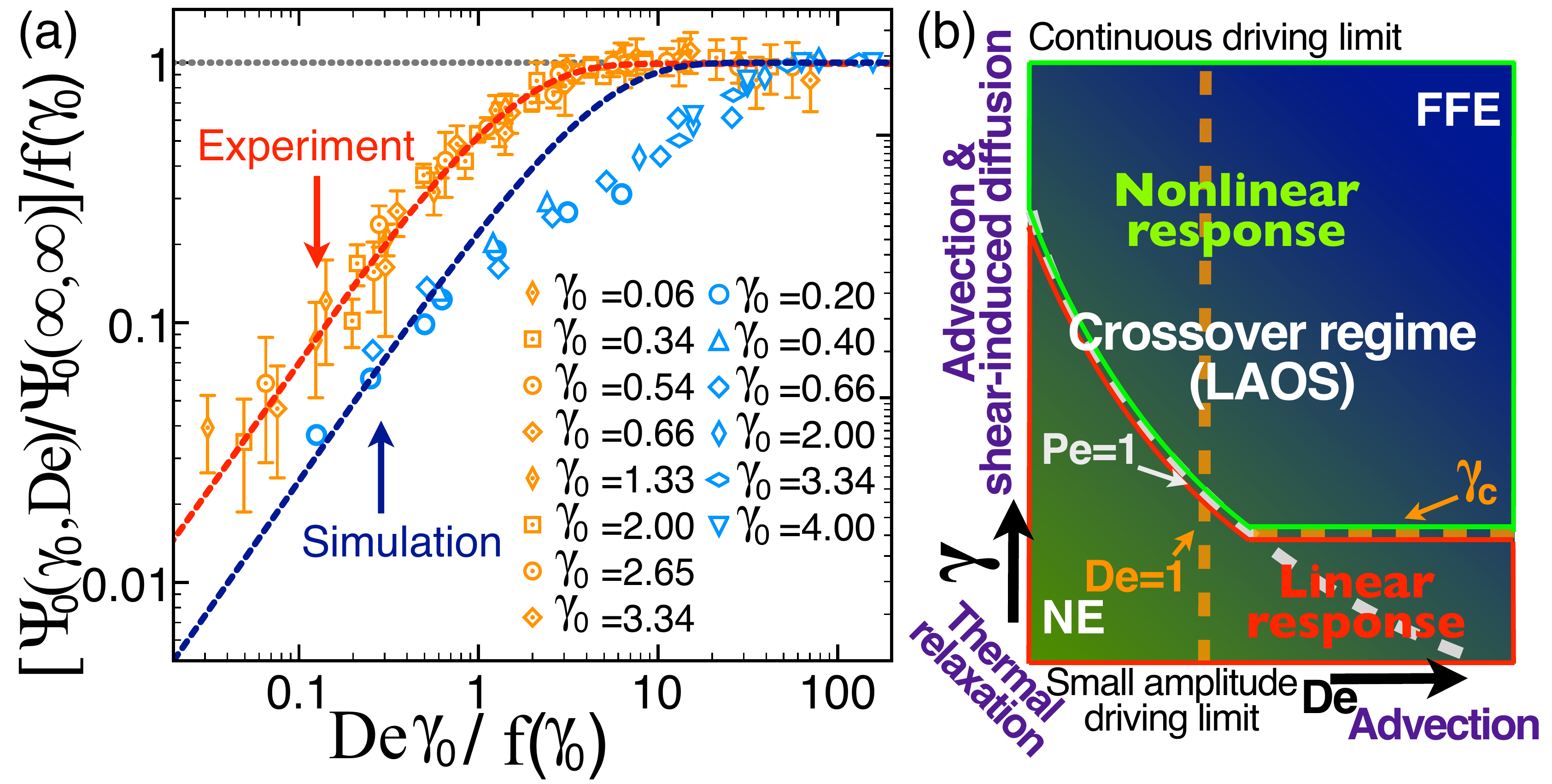}
\caption{(Color online) $\Psi_0(\gamma_0,{\rm De})$ collapse (a), and phase diagram of LAOS crossover (b). (a), $[\Psi_0(\gamma_0,{\rm De}) / \Psi_0(\infty,\infty)]/f(\gamma_0)$ versus the dimensionless scaling parameter De $\gamma_0 / f(\gamma_0)$, where $f(\gamma_0)= 1 - e^{-\gamma_0/\gamma_c}$. Each symbol denotes one strain amplitude for seven different De in the range $0.39<$ De $<18.8$. All 56 data points collapse on a master curve and can be fit by an exponential saturation (upper red line) $1-e^{-\beta \gamma_0 De/f(\gamma_0)}$, where $\beta=0.72$ is a fitting parameter. Error bars correspond to standard errors of the data averaged over five runs. (b) Three entangled dynamics (bold purple fonts) result in two types of response saturations. In the crossover regime bounded by nearly equilibrium (NE) and far-from-equilibrium states (FFE), nonlinear response emerges when Pe$>$1 and $\gamma>\gamma_c$.}
\label{fig:fig4}
\end{figure}

These saturation behaviors can be understood by considering the microscopic particle dynamics. In the experiments, $\Psi_0(\gamma_0,{\rm De})$ reflects the maximum degree to which particles accumulate along the MCA due to shear. The data trends for small strain amplitudes can be understood as resulting from competition between Brownian relaxation of particles to their equilibrium configuration and advection resulting from the suspending fluid as described by the advection-diffusion equation\cite{Bird_book}. Hydrodynamic intereactions can also affect details of the suspension microsctructure. However, in the low De regime these effects are weaker than those played by Browninan motion \cite{Bergenholtz}, and in the high De regime their effect on the shape of the distorted microstructure still results in only a weak quantitative difference in the $g(\vec{r})$ at contact \cite{Morris1997}. Thus we focus on the interplay between Brownian motion and advection.

In the low De regime, Brownian diffusion dominates but becomes less effective at homogenizing particles with increasing De. Consequently, $\Psi_0(\gamma_0,{\rm De})$ increases linearly with De as particles accumulate along the MCA. In the high De regime, oscillatory flow dominates and thermal relaxations are negligibly weak: particles are simply advected by the flow. Thus, the plateau value $\Psi_0(\gamma_0,\infty)$ is set by the extent to which particles can accumulate along the MCA for a given strain amplitude. Lastly, the  crossover between these limiting behaviors occurs at De $\approx 1$. 

The initial linear increase in $\Psi_0(\gamma_0,\infty)$ (Fig.~\ref{fig:fig3} (b)) reflects the fact that for small $\gamma_0$, where there are very few collisions with neighboring particles, particle accumulation along the MCA increases with $\gamma_0$. As $\gamma_0$ approaches $\gamma_c$, particles collide more frequently thereby randomizing particles accumulated along the MCA. The saturation of $\Psi_0(\gamma_0,\infty)$ for $\gamma_0 > \gamma_c$ indicates there exists a limit to which particles can be driven to accumulate along the MCA. This limit results from a competition between collision-induced randomization \cite{Pine2008, Pine2009, Foss1999, Sierou2004, Leighton1987} and advection. Thus, $\gamma_c$ corresponds to the average strain needed to induce collisions between neighboring particles.  

This argument suggests that for higher volume fractions where interparticle distances are smaller, $\gamma_c$ should be smaller. Specifically, we expect that at $\gamma_0 = \gamma_c$, the relative distance traveled by particles separated by a diameter along the gradient direction, $2a \gamma_0$, will equal the mean distance between neighboring particles. From geometrical considerations this distance scales as $a(\frac{\phi_c-\phi}{\phi})^{1/3}$, where $\phi_c = 0.64$ is the random close packed volume fraction. Thus, $\gamma_c \propto (\frac{0.64-\phi}{\phi})^{1/3}$ up to a constant. 

To test this prediction we measure $\Psi_0(\gamma_0,\infty)$ for six different volume fractions ranging between 0.15 to 0.44. For each curve we extract both $\gamma_c$ and $\Psi_0(\infty,\infty)$ and plot these values as a function of $\phi$ (Fig.~\ref{fig:fig3}(b) inset) along with the fit from our model (solid curve). We find excellent agreement between the $\gamma_c$ data and the model fit. We also find that for the range of $\phi$ values measured, $\Psi_0(\infty,\infty)$ increases monotonically with $\phi$. Finally, we find the data from eight different runs where $0.06 \leq \gamma_0 \leq 3.34$ and $0.09 \leq \rm De\leq 18.8$ collapses onto a curve of the form:
\begin{equation}
\frac{\Psi_0(\gamma_0,\rm De)}{\Psi_0(\infty,\infty)} = f(\gamma_0)\Big[1- e^{-  \gamma_0 {\rm De} \beta / f(\gamma_0)}\Big] \ \ 
\label{eq:scaling}
\end{equation}
%\begin{equation}
%\frac{\Psi_0(\gamma_0,\rm De)}{\Psi_0(\infty,\infty)} = f(\gamma_0) \Big\{ 1- \exp\big[ -\beta \gamma_0 {\rm De}/f(\gamma_0) \big] \Big\}\\
%\label{eq:scaling}
%\end{equation}
where $f(\gamma_0) = 1 - e^{-\gamma_0/\gamma_c}$ and $\beta$ is a fitting parameter (Fig.~\ref{fig:fig4}(a) red curve). This functional form recovers the linear regime, where the normalized structure response $\tilde{\Psi}=\Psi_0(\gamma_0,\rm De)/\Psi_0(\infty,\infty)$ is equal to $\gamma_0 {\rm De}$, when either De or $\gamma_0 \to 0$ ; the high-frequency regime $\tilde{\Psi} = f(\gamma_0)$, when De $ \to \infty$ ; and the high-strain amplitude limit $\tilde{\Psi} = 1$, when $\gamma_0 \to \infty$. More broadly, these results are summarized by the phase diagram in Fig.~\ref{fig:fig4}(b) that shows the crossover between near-equilibrium linear response and far-from-equilibrium nonlinear response. 
% In summary, these data indicate that the normalized structural response $\tilde{\Psi}$ is a function that depends on two separate dimensionless parameters, De, and $\gamma_0$. 
%Our results indicate that while the mechanism underlying each saturation is distinct, the curves collapse because their saturations have sufficiently similar functional forms.

\begin{figure} 
\includegraphics[width=0.48 \textwidth]{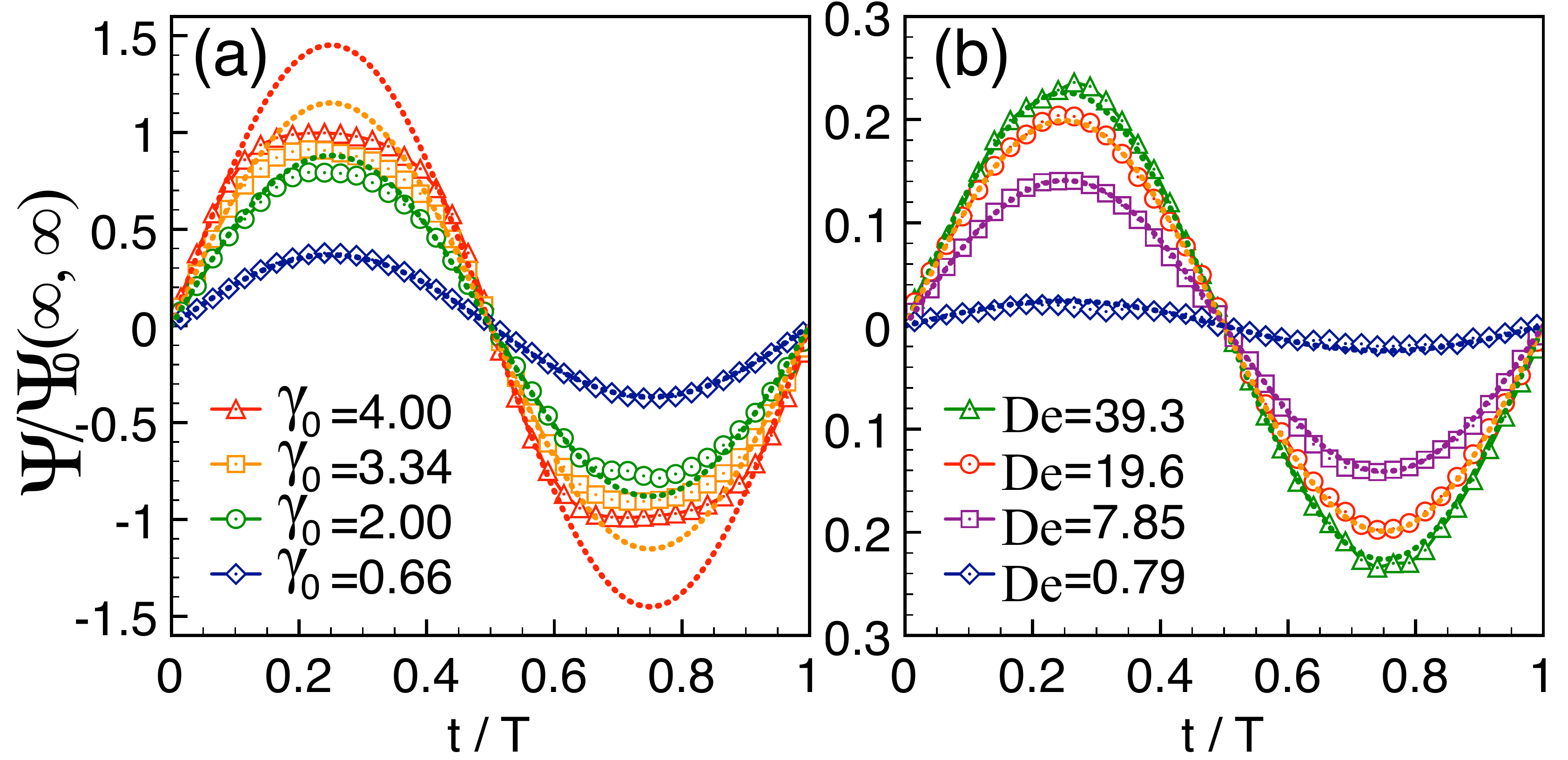}
\caption{(Color online) $\Psi_0(\gamma_0,{\rm De})$ in Brownian dynamics simulations. (a), Amplitude saturation. $\Psi_0(\gamma_0,{\rm De})$ is plotted for four different $\gamma_0$ at De = 15.7. (b), Frequency saturation. $\Psi_0(\gamma_0,{\rm De})$ is plotted for four different De at $\gamma_0$ = 0.20.}
\label{fig:simulation_figure_panel}
\end{figure}

\section{Simulation}

The interactions between hard sphere colloids leading to the observed saturations can be mediated either by collisions or hydrodynamics. To determine whether particle collisions are sufficient to generate such saturations, we conduct dynamic simulations using the LAMMPS package (Sandia National Laboratory). We implement Brownian Dynamics simulation by applying a Langevin thermostat to the streaming velocity of  simulated particles to maintain a constant temperature of T$^*$=1. The interparticle potential is taken to be $U/ (k_B T) = r^{-c}$. We found that the stress response for $c = 50$ agrees well with reported results for hard spheres. The simulation setup contains 10,000 particles with $\phi=0.28$. We apply oscillatory shear to the system with the Lees-Edwards boundary condition. Since the interparticle potential is very steep, the time step is carefully chosen to avoid unphysical particle overlaps. This model provides insights into the limiting physical behavior that ensues when pair- and higher-level hydrodynamic interactions are neglected. We run 100 oscillatory cycles for each De and $\gamma_0$, and discard data obtained from the first 10 cycles as transient. In direct analogy with the experiments, we calculate $\Psi(t)$ using Eq.~\ref{eq:model} with a radial integral that extends up to the first peak of $g(\vec{r})$ (see Appendix G).

We perform an amplitude sweep at De = 15.7 for six different amplitudes, and plot $\Psi(t)/\Psi_0(\infty,\infty)$ for four different $\gamma_0$ in Fig.~\ref{fig:simulation_figure_panel} (a). As with the experiments (Fig.~\ref{fig:result2} (a)), this model system also demonstrates an amplitude saturation at large $\gamma_0$, where $\Psi(t)/\Psi_0(\infty,\infty)$ deviates from the linear response. We also perform a frequency sweep for $0.20 \leq \gamma_0 \leq 4.00$ and plot the results for four of the six amplitudes in Fig.~\ref{fig:simulation_figure_panel}(b). We find that the model system exhibits a similar saturation to that found in experiments (Fig.~\ref{fig:result2}(b)). To determine whether similar data scaling can be applied to the numerical results, we plot the normalized value $\tilde{\Psi}/f(\gamma_0)$ versus $\gamma_0 {\rm De}/f(\gamma_0)$ in Fig.~\ref{fig:fig4}(a). We find that the simulation data also collapses, but the curve's form deviates from the experimental curve at intermediate shear rates. Nevertheless, the collapse is qualitatively similar to the experimental results showing a linear response at low shear rates and saturation at high shear rates. These results demonstrate that the interplay between Brownian relaxation, advection and shear induced diffusion is sufficient to produce the observed saturations. 

%%%%%%% Simulation ends%%%%%%%%%%

\section{Conclusion}

Measuring suspension structure is a noninvasive and explicit method to quantify each contribution of the macroscopic stress response\cite{Cheng2011}. Specifically, direct imaging allows one to measure the contribution due to Brownian motion of the microscopic constituents, which is a vital component to the response of any thermal system. Previous theoretical work showed that the surface integral of $\hat{r} \hat{r} g(\vec{r})$ at $r=2a$ is proportional to the pairwise Brownian stress\cite{Foss2000,Brady1993}. In experiments, optical resolution limits require that a radially averaged quantity is used instead (Eq.~\ref{eq:model}). This modified calculation has been shown to agree with macroscopic force measurements in the nearly equilibrium regime\cite{Cheng2011}. While it remains to be shown that this modified expression is a valid measure of the Brownian stress in the far-from-equilibrium regime, our results on the saturations of $\Psi_0$ are consistent with bulk rheological measurements reported previously\cite{Giuseppe2012,Brader2010}, suggesting a very strong link between $\Psi_0$ and the Brownian stress. 

Since the shear separation in the experiment is approximately seven particle diameters, the confinement effect may be significant in the reported system. Even though the particles that are near boundaries are excluded in analysis, the long-ranged hydrodynamic interactions between wall and particles may still play an important role in determining the dynamics and configuration of particles. Furthermore, the interparticle hydrodynamic interactions are left out in the Brownian dynamics simulation for identifying the origin of the structure response saturation. The fact that the Brownian dynamics simulations do not perfectly reproduce the experimental data indicate that HI do have some effect on the particle distributions. However, this effect is not large enough to qualitatively alter the trends. Namely we still observe linear scaling at low Pe and a saturation with higher frequency or strain amplitude, see Fig.~\ref{fig:fig4}(a) and the Appendix. Nevertheless, conducting full hydrodynamic simulation that accurately take into account the boundary conditions at the surface is necessary for having a more rigorous and quantitative comparison with our experimental results.

Many different complex systems such as emulsions\cite{Bousmina2006}, plasmas\cite{Nosenko2005}, and polymers\cite{Hyun2011} exhibit saturation behaviors when driven away from equilibrium. For example, polymer blends have the same measured viscosity whether driven by continuous shear (${\rm De} \ll 1, \gamma_0 \to \infty$) or  perturbed with small amplitudes at high frequency ({\rm De}$\to \infty$, $\gamma\ll1$).
Understanding the underpinnings of this well known but poorly understood empirical observation, known as the Cox-Merz rule\cite{Cox1958,Larson}, has remained a long standing theoretical challenge \cite{Doraiswamy1991,Hadithi1992}.
By combining direct imaging with LAOS, we find an analogous behavior in the double saturation of the suspension structure response. Here, we show this double saturation can be collapsed on a master curve (Eq.~\ref{eq:scaling}), which identifies the roles of Brownian relaxation, affine motion, and shear induced diffusion. In part, this finding is made possible by combining direct imaging with LAOS, a shear protocol that disentangles these dynamics (Fig~\ref{fig:fig4}(b)). Brownian dynamics simulations show that the interplay between these three elements is sufficient to generate the Cox-Merz rule, analogous to other driven far-from-equilibrium systems. Further experiments with these techniques should elucidate additional mechanisms in the crossover regimes between nearly equilibrium and far-from-equilibrium states, and shed light on the highly nonlinear dynamics found in many other far-from-equilibrium systems.

\section*{ACKNOWLEDGMENTS}

We thank D. Koch, L. Archer, I. Procaccia, G. Henchel, E. Bochbinder, B. Leahy, and J. L. Silverberg for helpful conversations. This work was supported in part by Award No. KUS-C1-018-02 made by King Abdullah University of Science and Technology (KAUST), the U.S. Department of Energy, Office of Basic Energy Sciences, Division of Materials Sciences and Engineering under Award No. ER46517 (X. C.) and the National Science Foundation CBET-PMP award No. 1232666. FAE is grateful to computer cycles supplied by the Extreme Science and Engineering Discovery Environment (XSEDE) which is supported by National Science Foundation grant number OCI-1053575.

\appendix

%%%%%%%% Sample characterization and imaging resolution %%%%%%%%%%%

\section{Sample characterization and imaging resolution}
\begin{figure} [htp]
\includegraphics[width=0.48 \textwidth]{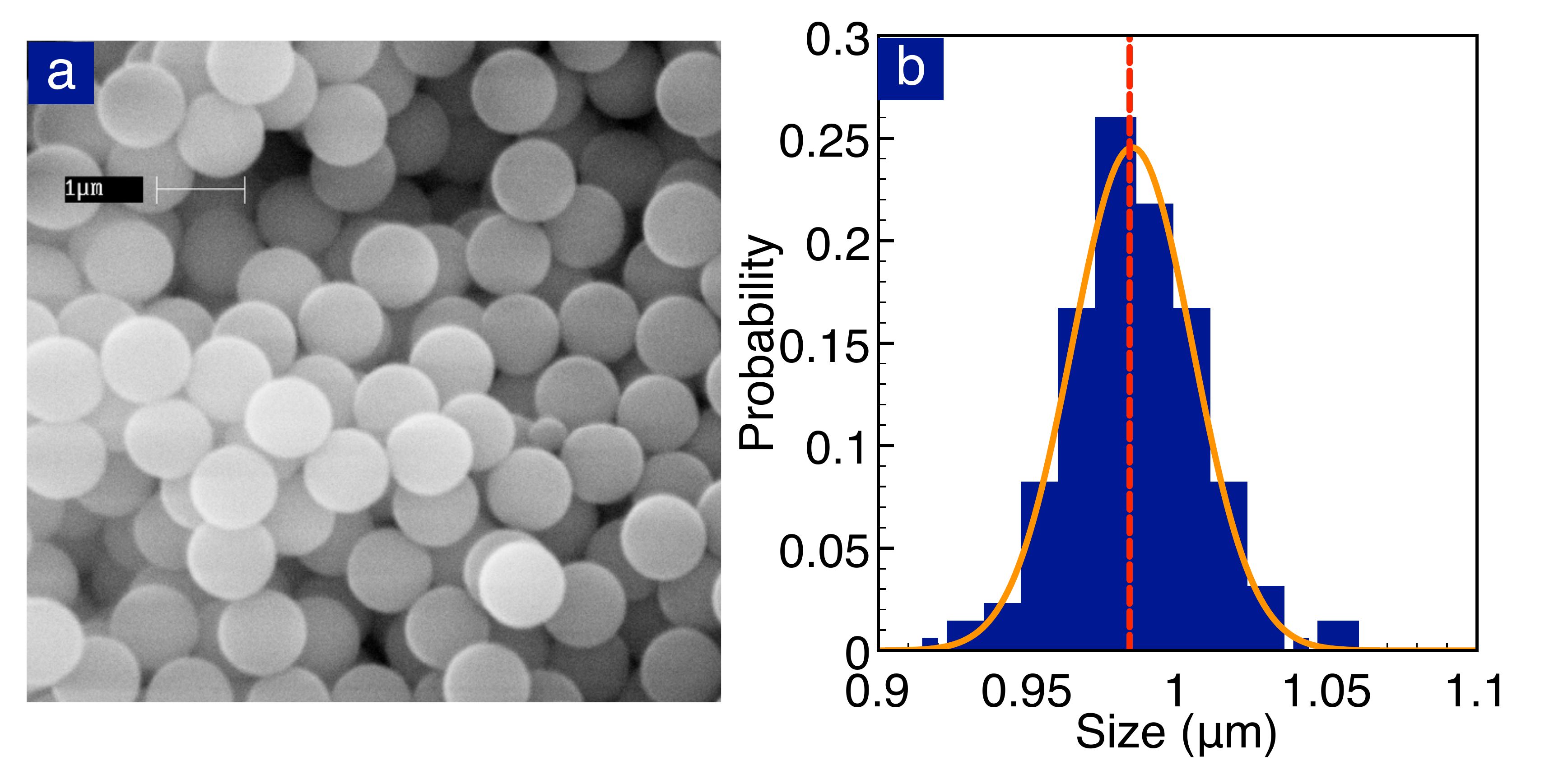}
\caption{(Color online) (a) SEM image of the silica particles. (b) The probability distribution function of the particle size. The orange solid line is a fit of the Gaussian distribution, and the red vertical dashed line at 0.98$\mu$m delineates the mean of the distribution.}
\label{size_distribution}
\end{figure}

To measure the size distribution of our silica particles, we acquired a series of SEM images (Leica 440 SEM) of the particles (Fig.~\ref{size_distribution} (a)). We measured the sphere size from the SEM images. Fig.~\ref{size_distribution} (b) shows the size distribution of the sample in Fig.~\ref{size_distribution} (a) based on the statistics of 200 particles. Because the particles are spherical and the screening length $\sim10$nm is short, the hydrodynamic radius and the hard sphere radius are nearly the same in this experiment. 

In Fig.~\ref{size_distribution} (b), we find that the distribution of the particle size is well fitted by a Gaussian distribution with a mean value $2a=980$nm, and a standard deviation $2 a\times2.05\%$. 

Our resolution for locating particles is about 50nm. Since the circumference of a particle is $2 \pi (2a)\approx 6,280$ nm, our angular resolution ends up being about three degrees. If all particles were to have neighbors along the maximal compression axis the exact result would give a $\Psi$ of 0.5. We estimate that a 50nm error would lead us to calculate a $\Psi$ of 0.499. Therefore, the measurement error due to our imaging resolution is not significant. 

%%%%%%%%%% Actual $g_{xy}$ plots %%%%%%%%%% 

\section{Raw data of $g_{xy}$}

\begin{figure} [htp]
\includegraphics[width=0.5 \textwidth]{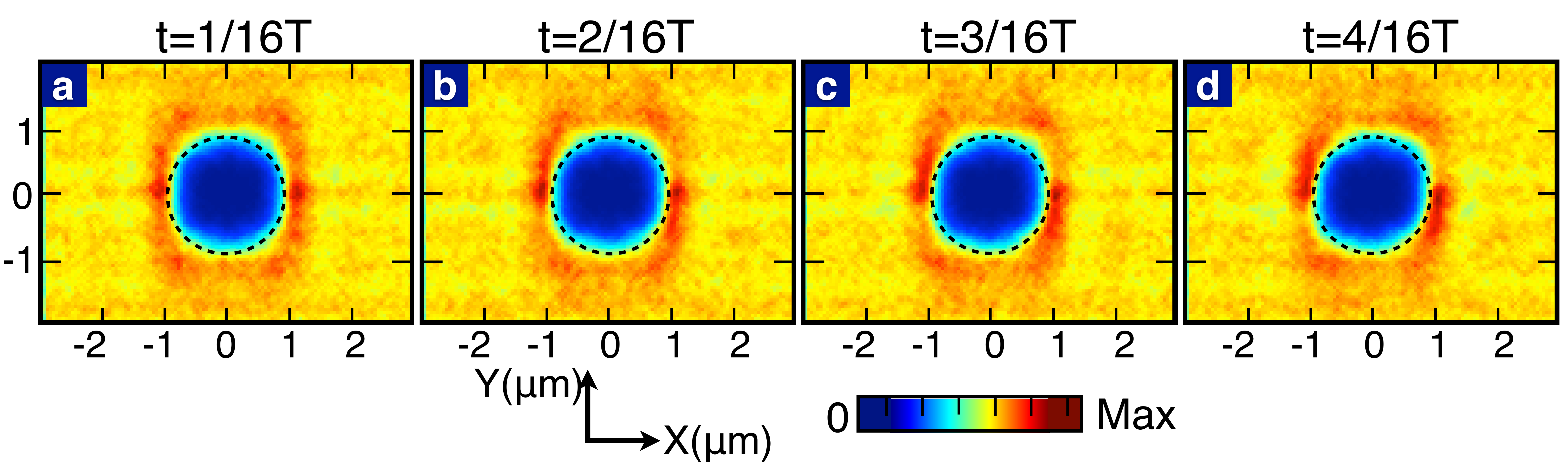}
\caption{(Color online) $g_{xy}$ without subtraction. The corresponding experimental parameters are the same as the ones of Fig.~1 in the main report.}
\label{nonsub}
\end{figure}

We plot the actual $g_{xy}$ in Fig.~\ref{nonsub} to illustrate the evolution of the sheared suspension structure. Fig.~\ref{nonsub} and Fig.~1 of the main report share the same data set. The squared first peak of $g_{xy}$ indicates that the system has layering due to the confinement effect. It is important to point out that the layer structure does not have any contribution to the value of $\Psi$ due to the mirror symmetry of $g_{xy}$ at $t=0$.

%%%%%%%% Radial integral range %%%%%%%%%%%

\section{The radial integral range}
Many methods have been used to quantify the structure response of systems under shear. For instance, the ellipticity has been used to illustrate the distortion of $g(\vec{r})$ in dusty plasmas\cite{Nosenko2005}, the bond order parameter $\Psi_6$ has been employed in colloidal crystal under shear\cite{Peng2010,Hernandez-Guzman2009}, and the alignment factor $\Delta A$ has been employed to determine the orientation of the assembly of particle strings\cite{vermant2010a, Maranzano2002}. Here, we specifically calculate the angular probability distribution of neighboring particles to characterize the anisotropy of $g(\vec r)$. This anisotropy of $g(\vec r)$, which is closely related to the stress response, quantifies the response of the microstructure to the external shear flow\cite{Brady1993}. To quantify the structural response, we integrate $g(\vec{r})$ up to its first peak to account for particle contacts.
We find that in our experiments the first peak of $g(\vec{r})$ is at $r = 2.35a$. Thus we define the shear or $XY$ component of our structural signature as:

\begin{equation}
\Psi_{B} = [\frac{1}{0.51a} \int_{1.84a}^{2.35a} dr \oint \hat{r} \hat{r} g(\vec{r}) d\Omega]_{XY}.
\label{eq:model}
\end{equation}

\begin{figure} [htp]
\includegraphics[width=0.47 \textwidth]{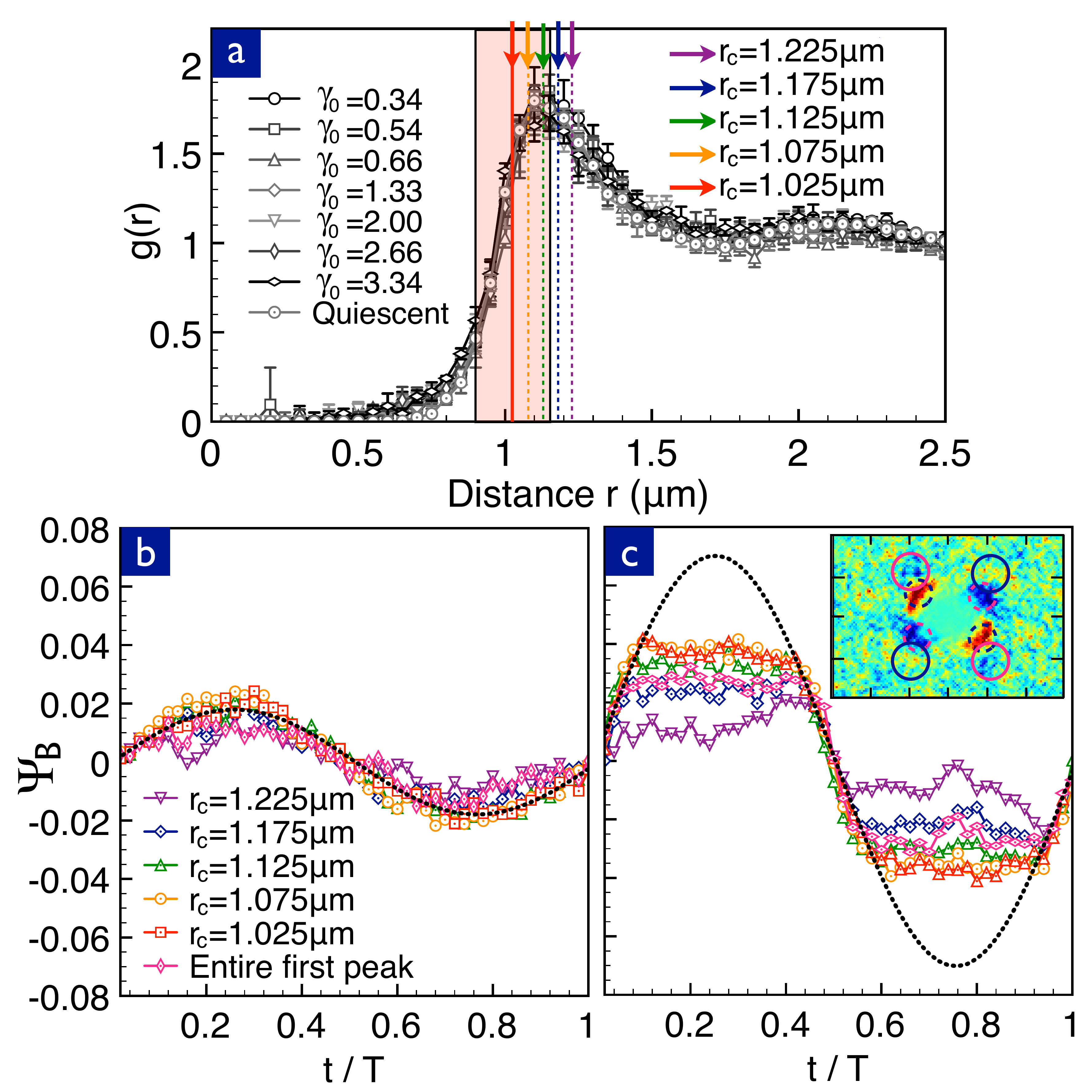}
\caption{(Color online) Radial distribution functions $g(r)$ (a) and $\Psi_B$ for different integral bounds at $(\gamma_0=0.34, \omega=0.126s^{-1})$ (b) and $(\gamma_0=3.34, \omega=0.126s^{-1})$ (c). (a) 1D radial distribution functions of the suspension for seven different $\gamma_0$ at $\omega=0.126s^{-1}$ and the quiescent sample are plotted versus distance. The five arrows indicate the centers of the integral bounds that are tested for the calculation of $\Psi_B$. The red shaded area illustrates the integral bound of $r_c$=1.025$\mu$m. The vertical arrows from left to right correspond to the integral centers at $r_c$($\mu$m)=1.025, 1.075, 1.125, 1.175 and 1.225, respectively. (b) The raw oscillatory data of $\Psi_B$ at $(\gamma_0=0.34, \omega=0.126s^{-1})$ is plotted versus t/T for five different $r_c$ and the integral bound $0.90\mu$m$<r<$1.50$\mu$m. (c)$\Psi_B$ at $(\gamma_0=3.34, \omega=0.126s^{-1})$ is plotted versus t/T for five different $r_c$ and the integral bound $0.90\mu$m$<r<$1.50$\mu$m. Both dashed lines in (b) and (c) are the linear fit from Fig.~2(a) of the main paper. In the inset to (c), the solid circles highlight the wake structure of the $\Delta g(\vec{r})$ measurement taken from Fig~2(d) of the main paper.}
\label{integral_range_combine}
\end{figure}

Imaging artifacts associated with particle tracking errors are removed by introducing a lower bound to the radial integral at $1.84a$.  We find that narrowing this integral width in order to have a stricter criteria for the contacting particles does not alter the qualitative trend of $\Psi_B$ but only introduces more noise due to poorer statistics.

Furthermore, we have studied the dependence of the structural signature $\Psi_B$ on the radial integral center $r_c$. We show angular averaged $g(\vec{r})$ for eight strain amplitudes $\gamma_0$ with a fixed shear frequency $\omega=0.126$s$^{-1}$ in Fig.~\ref{integral_range_combine} a. All radial distribution functions overlap on a single curve within the error bars. This overlap indicates that using a single fixed integral bound for all the experiments presented in this paper produces consistent calculation results. The short tail at $r<2$a$=0.98\mu$m is mainly due to the finite resolution and the point spread functions along the vertical-axis  (the $Y$-axis) of the confocal microscope, as well as the particle polydispersity. This relatively poorer resolution results in an uncertainty in the particle featuring process and introduce a small portion of unphysical particle overlaps. We set the lower bound of the integral at 0.92$\mu$m to exclude those overlapping particles. With this lower bound, which is about four standard deviations from the mean, at least 94$\%$ of the particle population is included.

We calculate $\Psi_B$ with five different integral centers for the two data sets that have $(\gamma_0=0.34, \omega=0.126s^{-1})$ ( Fig.~\ref{integral_range_combine}b) and $(\gamma_0=3.34, \omega=0.126s^{-1})$ (Fig.~\ref{integral_range_combine}c). These data sets correspond to the linear high frequency response and the high amplitude nonlinear response of $\Psi_B$. The corresponding positions of $r_c$ are labeled as vertical lines in Fig.~\ref{integral_range_combine}a and cover the range $1.025\mu$m$\leq r_c \leq 1.225\mu$m which encompasses the first peak of $g(r)$. We find that the qualitative trends in $\Psi_B$ are not sensitive to the choice of $r_c$ (Fig.~\ref{integral_range_combine}b,c). We have also verified that increasing the integral range to $0.90\mu$m$<r<1.50\mu$m does not alter the qualitative trends in $\Psi_B$ (entire first peak data). 

The particular choice of $r_c$ and the integral range do however, have a quantitative effect. For example, the deviation from the sinusoidal fit in Fig.~\ref{integral_range_combine}b increases with increasing $r_c$. For the nonlinear response, we find that as $r_c$ increases, the magnitude of $\Psi_B$ decreases. 
These smaller magnitudes result from the angular inhomogeneity in $\Delta g(\vec{r})$. As shown in the inset of Fig.~\ref{integral_range_combine}c, positive and negative regions are paired so that a larger integral range in $r$ leads to cancellations that result in smaller values of $\Psi_B$. This collision induced wake structure has also been reported in previous simulations \cite{Zia2013}. In conclusion, we find that the value of $\Psi_B$ is qualitatively insensitive to the radial integral bound, and captures the anisotropy of the angular distribution of contacting particles in both the high frequency linear and high amplitude nonlinear regimes.

%%%%%% Pipkin diagram %%%%%%%%%

\section{Pipkin diagram}
\begin{figure} [htp]
\includegraphics[width=0.48 \textwidth]{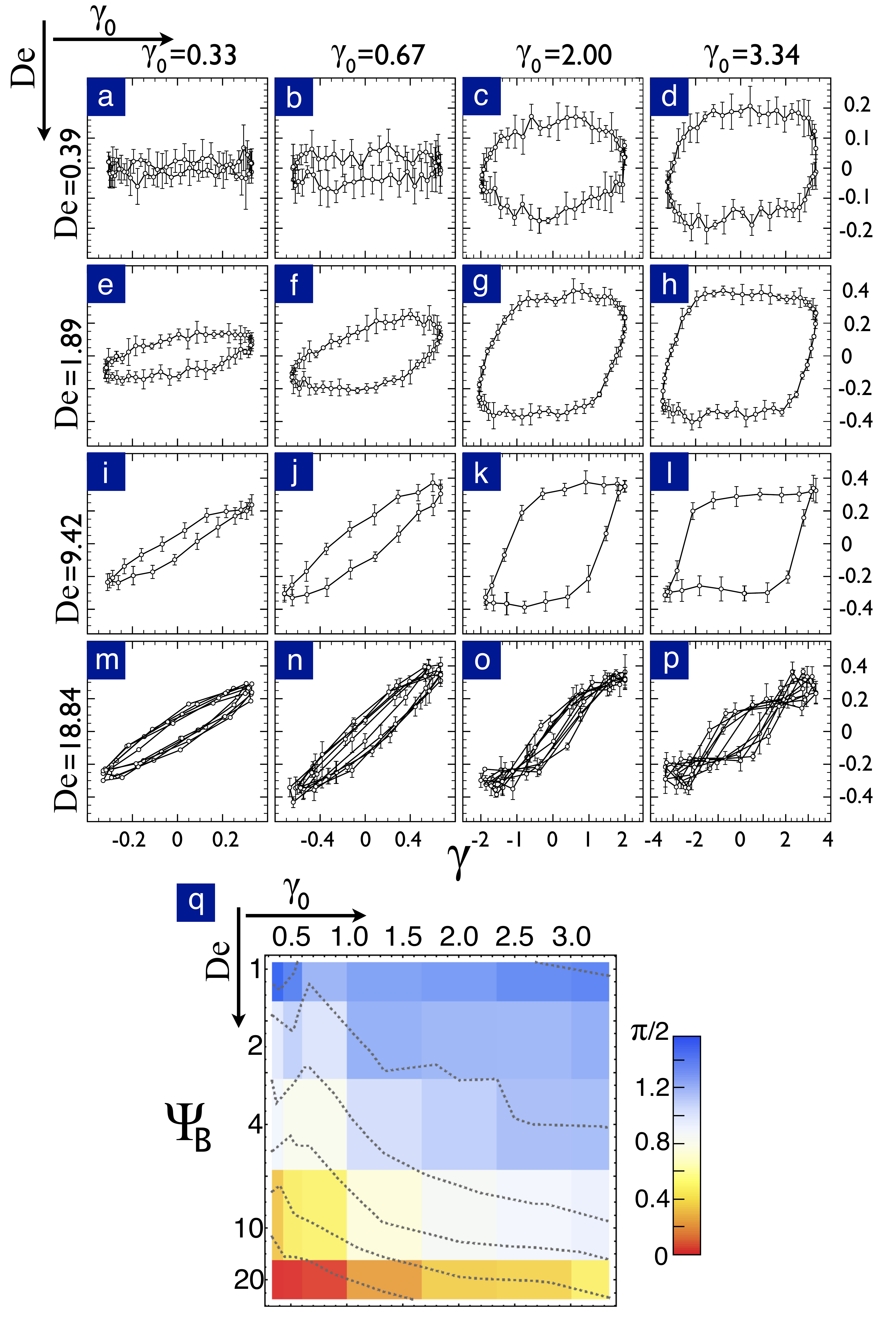}
\caption{(Color online) (a) to (p) L-B curves of $\Psi_B$ versus $\gamma$ for $4\times4$ different $\gamma_0$ and De on a Pipkin diagram. Each curve is averaged over five cycles of shear and the error bars denote the standard deviations. In the fourth row of plots ((m) to (p)), the data are averaged over 20 cycles of shear and five curves are displayed for clarity. (q) Phase angle between strain and stress is plotted versus De and $\gamma_0$. The color scale represents the value of the phase angle. The dashed lines depict the equal phase angel contours. The centers of the squares are the positions of the data points.}
\label{Pipkin_diagram_combine}
\end{figure}

For suspensions that are driven into their nonlinear response regime, the resulting stress response can depend on multiple experimental parameters. Pipkin diagrams in which a matrix of Lissajous-Bowditch (L-B) curves are organized into one figure have been widely used to aid in elucidating the dependence on multiple parameters \cite{Helgeson2007,Adams2009,Rogers2011,Ewoldt2010}. For the problem of colloids under shear the two relevant dimensionless parameters are $\gamma_0$ and De. As such we plot 16 $\Psi_B$ versus strain L-B curves in a Pipkin diagram in Fig.~\ref{Pipkin_diagram_combine}.

At low shear rates (small De and small $\gamma_0$), the suspension structure demonstrates a viscous response, where $\Psi_B$ peaks as $\gamma$ passes through zero. 
As De increases with a fixed small strain amplitude $\gamma_0=0.33$ (the first column of Fig.~\ref{Pipkin_diagram_combine}), the L-B curves become increasingly oblique indicating that the suspension structure response is more elastic. This viscous to elastic transition is reminiscent of the linear viscoelasticity observed in macroscopic rheological measurements \cite{Cheng2011, Lionberger1994, Shikata1994}, and the elastic plateau observed in simulations \cite{Brady1993}. With increasing $\gamma$ the L-B curves become more hysteric indicating a larger degree of viscous dissipation. These trends are summarized by a plot of the phase difference between the applied strain and $\Psi_B$ (Fig.~\ref{Pipkin_diagram_combine} (q). For the highest De measured we find that the L-B curves exhibit an overshooting behavior which is illustrated by the figure-eight curves in Fig.~\ref{Pipkin_diagram_combine} (l), (o) and (p). Similar overshoots have been found in the rheological measurements with much denser suspensions. In colloidal glasses, the overshoot is associated with the cage breaking\cite{Koumakis2012,Koumakis2012a, Agarwal2011} while in colloidal crystals the overshoot is suggested to be related to the zig-zag relative motion between two layers of lattices\cite{Chen1994, Elliot1997, Ackerson1986, Lopez-Barron2012}. Whether similar mechanisms can explain the overshoots in our data for low volume fraction $\phi=0.28$ suspensions under LAOS remains unknown. Overall, these trends are very different from those exhibited by Maxwell materials where an increase in strain produces a more elastic response and highlight the unique properties of suspensions under LAOS.

%%%%%%%% Quiescent $g(r)$ in experiments and simulations %%%%%%%%%

\section{Quiescent $g(r)$ in experiments and simulations}
\begin{figure} [htp]
\includegraphics[width=0.32 \textwidth]{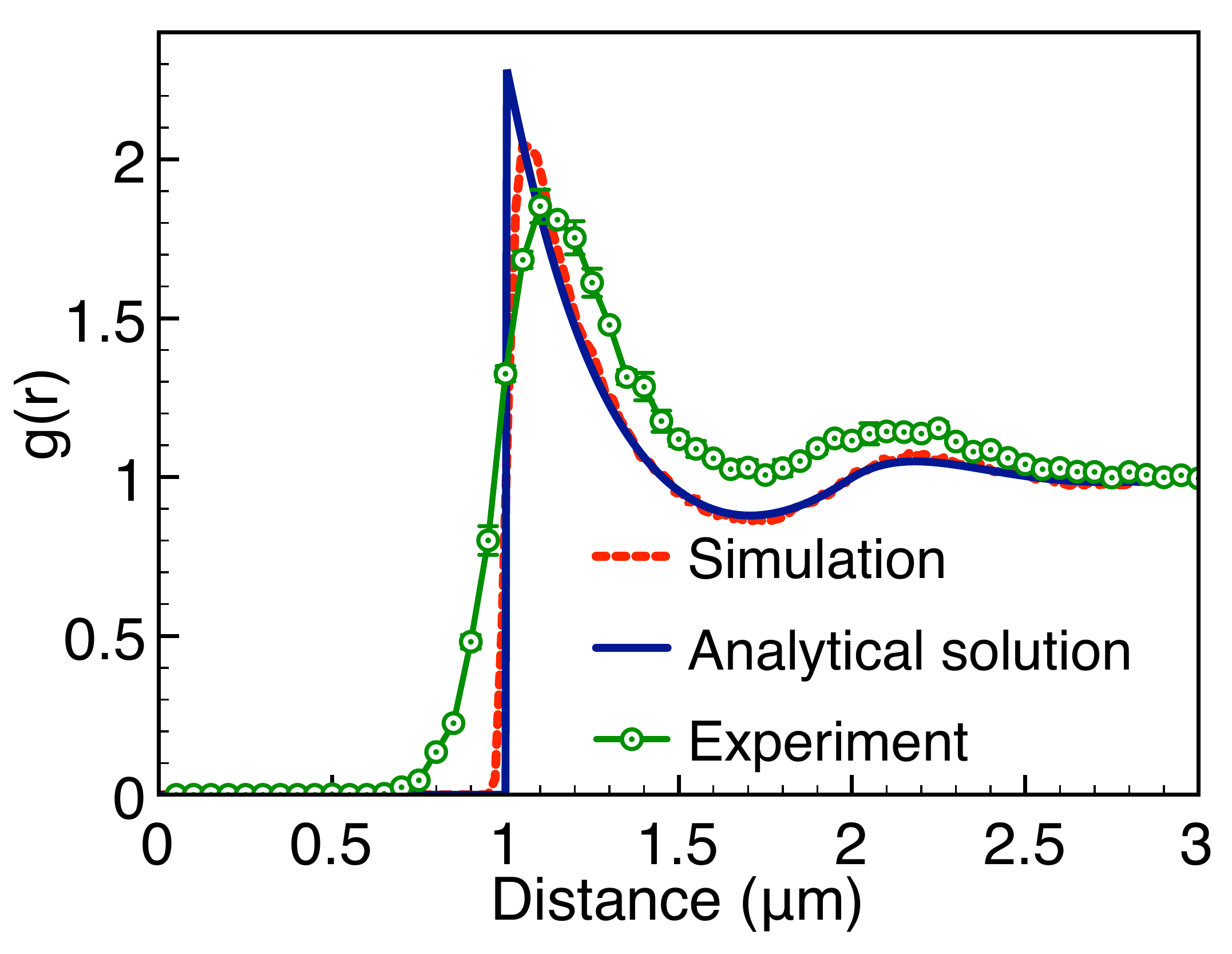}
\caption{(Color online) Comparison between the experimental (joined points), simulation (dashed line), and theoretical (solid line) $g(r)$ curves for quiescent samples of hard-sphere suspensions for a volume fraction of 0.28.}
\label{gr_exp_simulation}
\end{figure}

We compare the pair correlation functions $g(r)$ of static samples from the Brownian Dynamics simulation, analytical calculation, and the experiment in Fig.~\ref{gr_exp_simulation}. We find that all the curves are qualitatively similar and that the simulation and analytical results agree quantitatively. The analytical result is calculated from Percus-Yevick Integral Equation\cite{Trokhymchuk2005, Wertheim1963, Yuste1991}. 
The good agreement between the simulation curve and the theoretical prediction indicates that the steep potential used in the simulations is a good approximation to the hard sphere potential. The experimental $g(r)$ has a longer extension into the overlapping region ($r<1\mu m$) while the $g(r)$ from the simulation has a sharp drop at $r=1\mu$m. To understand this discrepancy, we also simulate the $g(r)$ with the polydispersity of our sample ($2\%$). We find that the $g(r)$ with $2\%$ polydispersity is indistinguishable from the curve for perfectly monodisperse spheres (not shown). This comparison implies that the tail extended in the overlapping region is due to the particle featuring errors. The magnitude of the particle featuring errors is set by limitations in imaging resolution, mismatched index of refraction, and shape of the point spread function.

\section{Quantification of the nonlinearity}

In addition to reporting on the peak value of $\Psi_B$ we further analyze the data to quantify the saturation at large $\gamma_0$ and De. There are a number of generic methods currently being used to quantify nonlinearities in the stress response of materials under LAOS \cite{Ewoldt2008, Hyun2011}. However, these methods require a large amount of data to perform an accurate measurement of the higher order harmonics. In direct imaging experiments however, each experiment run only acquires data for five oscillation cycles 
since each data point entails scanning the sample in 3D, tracking the particles positions, and calculating the variations in the pair correlation function. As such the data yield is not sufficient for running a harmonic analysis. Nevertheless, since the majority of the nonlinear response is associated with saturation plateaus, we can calculate the deviation of $\Psi_B$ from the linear response \cite{Emancipator1993} by defining the degree of nonlinearity $\lambda$ as:

\begin{multline}
\lambda=\frac{1}{T} \Bigl \{ \int_{t_0}^{t_0+T/2} \bigl [ \Psi_l(t) - \Psi_B(t)\bigr ]dt \\
+ \int_{t_0+T/2}^{t_0+T}\bigl [\Psi_B(t)-\Psi_l(t) \bigr ]dt \Bigr \}
\label{eq:nonlinear}
\end{multline}

\begin{figure} [htp]
\includegraphics[width=0.48 \textwidth]{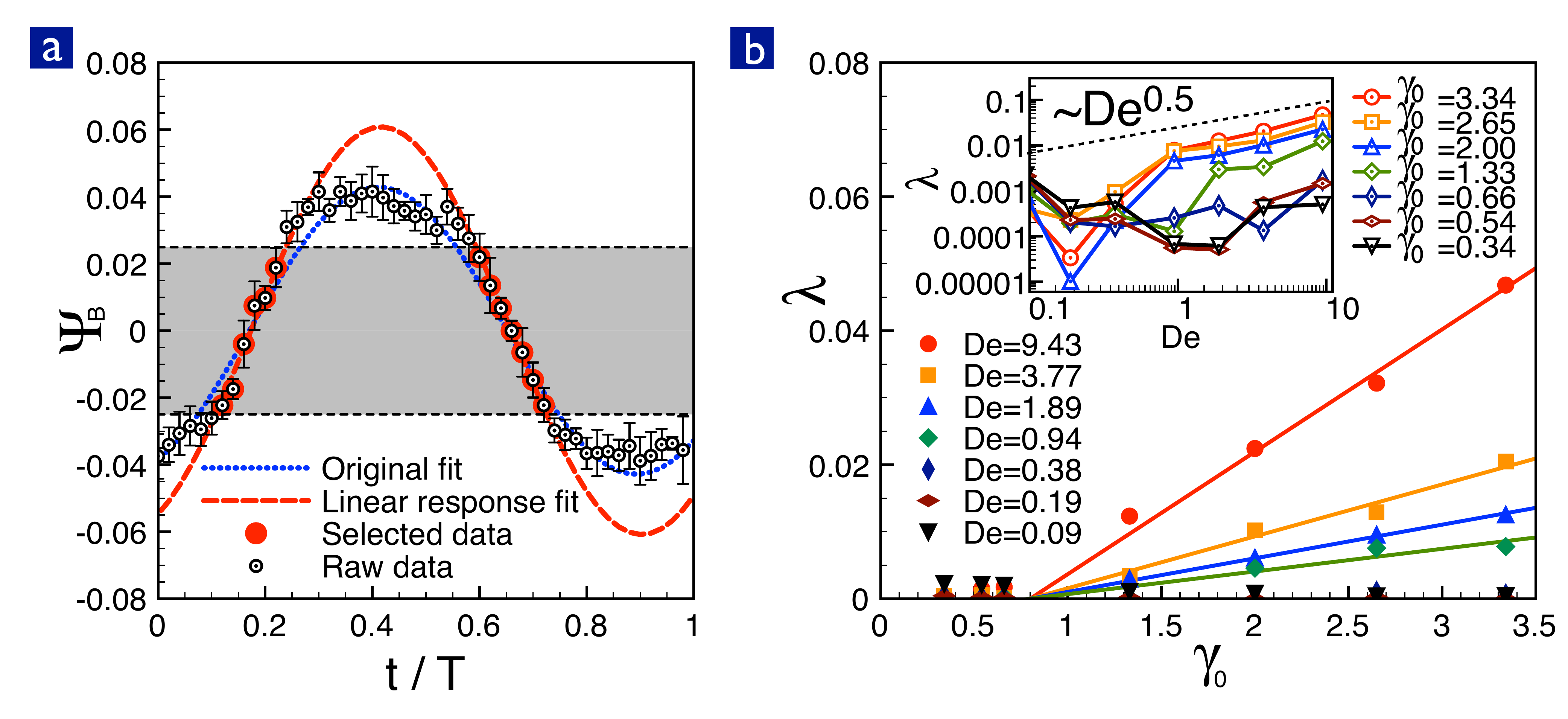}
\caption{(Color online) Definition of the degree of nonlinearity $\lambda$ (a), and $\lambda$ as a function of $\gamma_0$ for different De, (b). (a) The raw $\Psi_B$ data is plotted along with the best sinusoid fit to the raw data (blue dotted line), and the recovered linear response fit (red dashed line). The selected data ($\Psi_B\leq0.25$, gray shaded area) for the linear response fitting are highlighted by red circles. (b) $\lambda$ is plotted as a function of $\gamma_0$ for seven different De. The solid lines are the linear fits to the data for $\gamma_0>$ 0.75. The inset shows $\lambda$ versus De for different $\gamma_0$, and the dashed line that illustrates the power law $\lambda\sim$De$^{0.5}$.}
\label{nonlinearity}
\end{figure}

where $T$ is the period of one shear cycle, $t_0$ corresponds to the time where $\Psi_B = 0$ and $\Psi_B^\prime >0$, and $\Psi_l(t)$ is the recovered linear response. $\Psi_l$ is determined by fitting a sine wave to linear portion of the data where $\Psi_B \leq 0.025$. Eq.~\ref{eq:nonlinear} measures the deviation of the actual response, $\Psi_B(t)$, from the ideal linear response $\Psi_l(t)$. Integrating $|\Psi_l(t) - \Psi_B(t)\bigr|$ would produce integration artifacts due to noise. Instead, we divide the integral into two terms each of which integrates over a half cycle of the oscillation. This modified analysis significantly reduces the unbiased error by canceling random fluctuations.  To illustrate the difference between the recovered linear response and the original fit, we plot $\Psi_l$ (red dashed line) and the best fit to the raw data (blue dashed line) in Fig.~\ref{nonlinearity}a.

We calculate $\lambda$ and plot it as a function of strain amplitude $\gamma_0$ for seven different De in Fig.~\ref{nonlinearity} b. We find that for $\gamma_0 \leq 0.75$, the structure response $\Psi_B$ is linear at all De. This strain value is close to $\gamma_c = 0.24$ the cutoff strain obtained by fitting $\Psi_{De_\infty}$ to an exponential saturation $\Psi_{\omega_{\infty}}^{\gamma_{ \infty}}(1 - e^{-\gamma_0/\gamma_c})$. While we do not expect these two values to be identical, we do expect that the saturation behavior is closely tied to the nonlinearity and as such expect that $\gamma_0$ should be on the same order as $\gamma_c$. 

At large $\gamma_0$ and De, $\lambda$ increases linearly with $\gamma_0$. To study the De-dependence, $\lambda$ is plotted versus De for seven different $\gamma_0$ values in the inset of Fig.~\ref{nonlinearity} b. We find that $\lambda < 0.001$ and independent of De at small $\gamma_0$. At large $\gamma_0$ and De, we find that $\lambda$ is consistent with the form De$^{0.5}$. Whether this functional dependence extends to larger $\gamma_0$ and De, and what implications this functional form suggests are not known.

%\section{Raw data of simulation}
%We plot the raw simulation data for for four different $\gamma_0$ (see Fig.~\ref{simulation_raw} (a)) and four of the six amplitudes (see Fig.~\ref{simulation_raw} (a)). We find quantitative differences between the experimental and simulation results. For instance, the actual values of simulation data are approximately three times higher than the experimental data. In addition, the scaled simulation data deviate from collapsed experimental data curve in the intermediate $De \gamma_0/f(\gamma_0)$ regime as shown in Fig.~4 of main report. These quantitative discrepancies may be due to the differences between the experimental system and the Brownian dynamics simulations. In the experiment, all sheared particles collide and interact through hydrodynamically mediated lubrication forces, while in simulation all particles interact through interparticle forces. Furthermore, in experiments the system has approximately seven particles between shearing boundaries, while in simulation the system has a periodic boundary condition. 

%\begin{figure} [htp]
%\includegraphics[width=0.45 \textwidth]{simulation_raw}
%\caption{Raw oscillation data of the simulation results. The corresponding parameters are identical to the ones of Fig.~5 in the main report.}
%\label{simulation_raw}
%\end{figure}

\section{The effect of the interparticle potential on simulations}

\begin{figure} [htp]
\includegraphics[width=0.45 \textwidth]{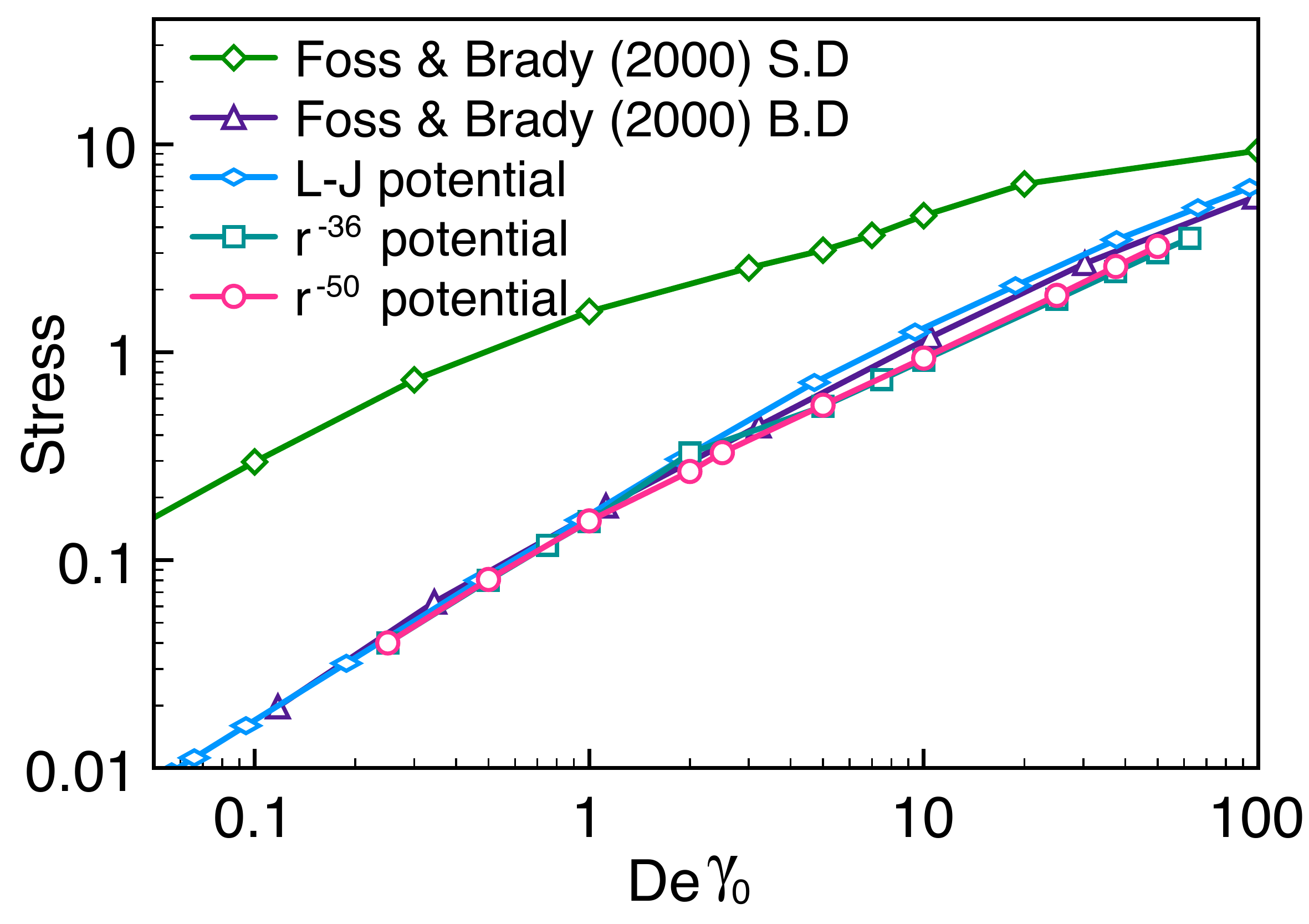}
\caption{(Color online) Stress responses in different simulations are plotted versus De$\gamma_0$. The data points of Foss \& Brady (2000) S.D and Foss \& Brady (2000) B.D are the Stokesian dynamics simulation ($\phi=0.45$) and Brownian dynamics simulation ($\phi=0.30$) from \cite{Foss2000} respectively.}
\label{diff_simulations}
\end{figure}

To examine whether the interparticle potential in Brownian dynamics simulations effectively mimics the hard sphere potential, we calculate the shear stress $\Sigma_{XY}$ for three different potentials, which are Lennard-Jones potential, $U \propto r^{-36}$ and $U \propto r^{-50}$. The results of all three different potentials are plotted along with the data reproduced from previous simulations results \cite{Foss2000} in Fig.~\ref{diff_simulations}. For the reproduced data, we determine the corresponding stresses by multiplying the $\eta$ value plotted in the reference by Pe/($6\pi r^3$). 

To examine whether the interparticle potential in Brownian dynamics simulations effectively mimics the hard sphere potential, we calculate the shear stress $\Sigma_{XY}$ for three different potentials, namely, the WCA Lennard-Jones-based (LJ) potential, $U \propto r^{-36}$ and $U \propto r^{-50}$. Fig.~\ref{diff_simulations} shows the results of all three different potentials are plotted along with data from previously reported simulations results for hard spheres \cite{Foss2000} (for the latter we determine the corresponding stresses by multiplying the reported $\eta$ value by Pe/($6\pi r^3$)). 

We find that when the steepness of the repulsive branch of the potential is increased from WCA Lennard-Jones potential ($U\propto r^{-12}$) to $U\propto r^{-36}$, the stress response is quantitatively similar at small De$\gamma_0$ and deviates very slightly at De$\gamma_0>10$ . As the potential steepness is increased from $U\propto r^{-36}$ to $U\propto r^{-50}$, the stress remains quantitatively similar at all De$\gamma_0$. This shows that the $U\propto r^{-36}$ and $U\propto r^{-50}$ potentials can both be considered very good approximations to  the hard sphere potential. The stress outputs from both potentials also nearly match the resulting stress from previous Brownian dynamics simulations, which use a different algorithm to generate the hard sphere potential\cite{Foss2000}. Results from Stokesian dynamics simulations for hard sphere suspensions are also available but for a packing fraction of 0.45\cite{Foss2000}, which is larger than that the 0.30 used in the Brownian dynamics simulations. Under such conditions, the calculated stress from Stokesian dynamics is approximately one order of magnitude larger than that found for the Brownian dynamics simulations at small De$\gamma_0$. At large De$\gamma_0$, the stress from Stokesian dynamics simulations appears to saturate while the stress in Brownian dynamics simulation keeps increasing with nearly constant rate\cite{Foss2000}.

%%%% Psi_B V.S Simga_xy %%%%%%%

\section{$\Psi_B$ versus $\Sigma_{XY}$ in Brownian dynamics simulation} 

In previous experimental and theoretical works, it has been shown that as the suspension is sheared, distortions of $g(\vec r)$ increase and lead to the Brownian stresses that arise from the thermal motion of particles\cite{Brady1993, Foss2000, Cheng2011}. Although the relative contribution from Brownian stress does decrease with increasing shear rate, it is important to note that this relative decrease arises because the hydrodynamic contribution grows linearly with strain rate while the Brownian contribution grows at a slower rate in this regime.

\begin{figure} [htp]
\includegraphics[width=0.45 \textwidth]{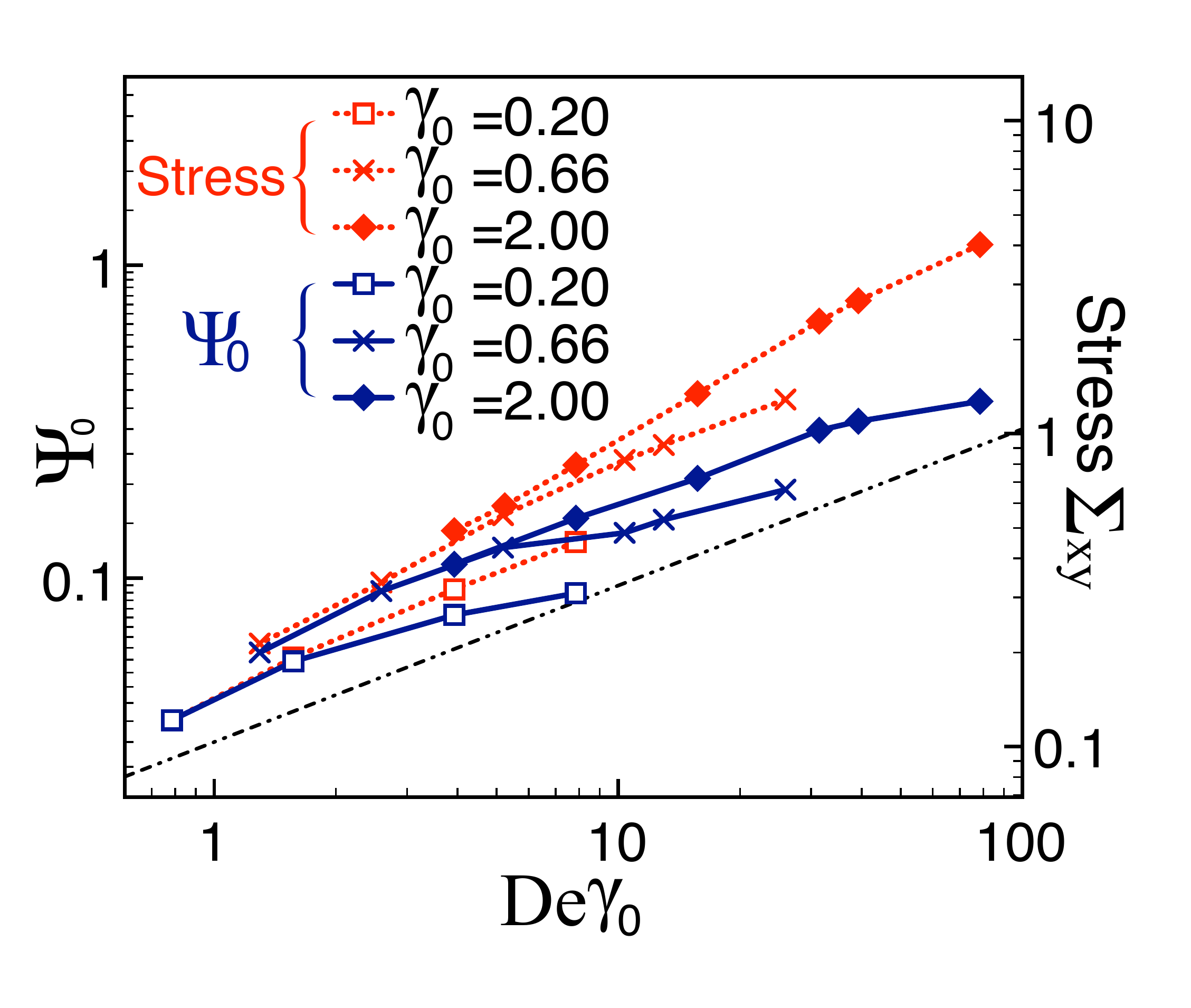}
\caption{(Color online) $\Psi_0$ and shear stress $\Sigma_{XY}$ are plotted versus De$\gamma_0$ for three different $\gamma_0$. The black dashed (straight) line illustrates the power-law $(De \gamma_0)^{0.5}$.}
\label{simulation_stress}
\end{figure}

To examine whether $\Psi_0$ reproduces the stress response in Brownian dynamics simulations, we directly calculate the stress tensor $\Sigma = \langle \vec{X}\vec{F}\rangle$, where $\vec{X}$ is the center to center position vector and $\vec{F}$ is the interparticle force. We plot $\Psi_0$ (solid lines) and the $XY$ component of the stress tensor $\Sigma_{XY}$ (dashed lines) as a function of De for three different $\gamma_0$ in Fig.~\ref{simulation_stress}. Here, we find good agreement for $\Psi_0$ and $\Sigma_{XY}$ at low shear rates, $\rm{De} \gamma_0\le5$, which is consistent with the previous experimental finding \cite{Cheng2011}. The values of the two calculations deviate from one another as De$\gamma_0>$5. This deviation between $\hat r \hat r g(\vec{r})$ integral and the stress response may be caused by the extra radial integral in the main manuscript Eq.~1. For example, in the high shear rate regime, the separation between particles becomes really small due to the strong shear flow. This narrow gap along with the divergent nature of the interparticle force may lead to large deviations between $\Psi_0$ and $\Sigma_{XY}$.
% Despite of the quantitative difference between two calculation, $\Psi_B$ captures the major shear thinning trend of the stress $\langle \vec{X}\vec{F}\rangle$ in a hard sphere system.

We also find that the trend of $\Sigma_{XY}$ is strikingly distinct from that of the existing rheological measurements \cite{Giuseppe2012, Brader2010} and our structure measurements. While $\Sigma_{XY}$ keeps increasing as De$\gamma_0$ increases, the latter two types of experiments show clear saturations at high De$\gamma_0$. This difference may indicate the increasingly important role of hydrodynamic interactions for colloidal suspensions driven away from equilibrium by LAOS. 

In contrast to the interparticle interactions in the Brownian Dynamics simulations, the Brownian stress in colloidal suspensions is transmitted through the solvent between particles. To calculate these hydrodynamically mediated forces, it is necessary to consider the mobility tensors, which require knowledge of the particle positions and velocities. It has been shown that as the separation between particles $\xi \rightarrow 0$, the mobility tensors are singular diverging as $1/\xi$. In addition, it was shown that because the singular force between particles is localized to the point of contact the final force calculation can be simplified to $n k_B T a \oint_{r=2a} \hat{r} \hat{r} g(\vec{r}) dS$, which has no $\xi$ dependence or particle relative position dependence. 

In our experiments, we find that the angular part of the pair correlation function, $g(\theta,\phi)$, has negligible dependence on the distance over the integral range. This observation suggests that one is able to perform a separation of variables on $g(\vec r)=g(r) g(\theta, \phi)$. Thus, Eq.~1 of main paper can be rewritten as:

\begin{equation}
\Psi_{B} = \frac{1}{0.51a} \Big(\int_{1.84a}^{2.35a} g(r) dr\Big) \Big( \oint_{r=2a} \hat{r} \hat{r} g(\theta, \phi) d\Omega \Big).
\label{eq:model2}
\end{equation}

The value of the first radial integral $\int g(r) dr$ is nearly constant at all shear rates as shown in Fig.~6a of main paper. The second integral, which only accounts for the contacting particles, is mathematically proportional to the Brownian stress $n k_B T a \oint_{r=2a} \hat{r} \hat{r} g(\vec{r}) dS$. Taken together, these results suggest $\Psi_B$ is proportional to the Brownian stress response of colloidal suspensions under LAOS.

To further examine whether $\Psi_B$ reports on the stresses it will be necessary to elucidate the role played by hydrodynamically mediated particle interactions. As such full hydrodynamic simulations will be necessary to more rigorously investigate the effect from the extra radial integral in Eq.~1 of main paper and the accuracy of estimating the total Brownian stress by only considering the pairwise term. Experimental confirmation of the ability to use $\Psi_B$ to report on the stress, will require precision force measurements that will allow for such comparisons.

%%%%%%%%%%%%%%%%%%%%%%%%%%%%%%%%%%%%%%%%%%%%%%%%%%%%%%%%%%%%%%%%%%%%%%%%%%%%%%%%
%%%%%%%%%%%%%%%%%%%%%%%%%%%%%%%%%%%%%%%%%%%%%%%%%%%%%%%%%%%%%%%%%%%%%%%%%%%%%%%%
%%%%%%%%%%%%%%%%%%%%%%%%%%%%%%%%%%%%%%%%%%%%%%%%%%%%%%%%%%%%%%%%%%%%%%%%%%%%%%%%
\end{document}